\documentclass[12pt]{article}
\usepackage{graphicx}
\usepackage{amsmath}
\usepackage{amssymb}
\usepackage[top=2.75cm, bottom=2.75cm, left=2.75cm, right=2.75cm]{geometry} 

\newcommand{\beq}{\begin{equation}}
\newcommand{\eeq}{\end{equation}}
\newcommand{\bea}{\begin{eqnarray}}
\newcommand{\eea}{\end{eqnarray}}
\newcommand{\om}{\omega}
\begin{document}
\begin{titlepage}
\begin{flushright}
YITP-SB-06-08\\
ITFA-2006-16\\
\end{flushright}
\vskip 22mm
\begin{center}
{\Large{\bf A Gravitational Effective Action on a\\[5mm] Finite Triangulation as a
Discrete\\[5mm] Model of Continuous Concepts\footnote{Presented at the
26th Winter School GEOMETRY AND PHYSICS at Srni by MR.}}}
\vskip 10mm
{\bf Albert Ko}\\
{\em Ward Melville High School}
\vskip 6mm
{\bf Martin Ro\v{c}ek}\\
{\em C.N. Yang Institute for Theoretical Physics}\\
{\em SUNY, Stony Brook, NY 11794-3840, USA}\\ 
and\\
{\em Institute for Theoretical Physics}\\
{\em University of Amsterdam, 1018 XE Amsterdam, The Netherlands}\\
{\tt rocek@insti.physics.sunysb.edu}
\end{center}
\vskip .2in

\begin{center} {\bf ABSTRACT } \end{center}
\begin{quotation}\noindent
We recall how the Gauss-Bonnet theorem can be interpreted as a finite
dimensional index theorem. We describe the construction given in 
{\tt hep-th/0512293} of a function that can be interpreted as a 
gravitational effective action on a triangulation. The
variation of this function under local rescalings of the edge lengths 
sharing a vertex is the Euler density, and we use it to 
illustrate how continuous concepts can have natural discrete analogs.
\end{quotation}
\vfill
\end{titlepage}
\newpage
\section{Introduction}
We want to study how we can translate concepts from continuum quantum field
theories to discrete models with a finite number of degrees of freedom. The particular
system that we explore is the theory of triangulations of a surface. This is a context that
is well understood in both the discrete \cite{regge} and the continuous cases \cite{poly}.

A nice example of such a translation is provided by the interpretation of the Euler
character as an index. In the continuous case, we have the exterior derivative:
\beq
d:~~\om_0\to\om_1\to\om_2~,
\eeq
where $\om_p$ are $p$-forms,
and the dual operator
\beq
*d*:~~\om_2\to\om_1\to\om_0~.
\eeq
We now consider the operator $D\equiv d+*d*$ restricted to
\beq
D:~~\om_0\oplus\om_2\to\om_1
\eeq
as well as its adjoint
\beq
D^\dagger:~~\om_1\to\om_0\oplus\om_2~.
\eeq
The index of $D$ is defined to be difference of the dimensions of the kernels of
$D$ and $D^\dagger$, and is known to be proportional to the Euler character $\chi$
of the surface \cite{duff,poly}:
\beq
dim(Ker(D))-dim(Ker(D^\dagger))\propto \chi=2(1-g)~.
\eeq

The discrete version of this well known story is somewhat 
less familiar. We consider a triangulation with $V$ vertices $v_i$, $E$
oriented edges $e_{ij}$, and $F$ oriented faces $f_{ijk}$. 
The discrete analogs of $p$-forms $\om_p$ are elements 
of a vector space $E_p$, where $E_0$ is associated to the 
vertices, $E_1$ is associated to the edges, and $E_2$ is associated 
to the faces. The sign associated to a edge or face 
depends on the orientation. There is an obvious notion 
of the exterior derivative $d$ which obeys $d^2=0$; 
the dual operator in general will depend on a choice of metric as
described in \cite{wilson}, but since the index is topological, we 
can ignore this dependence. Thus we choose our operators $D$ and 
$D^\dagger$ as follows:
\beq
(D\om)_{ij}=(\om_i-\om_j)\oplus(\om_{ijk}-\om_{ijk'})~,
\eeq
on an oriented edge $e_{ij}$ between a vertex $v_i$ and a vertex $v_j$ or on an
edge shared by oriented triangles $f_{ijk}$ and $f_{ijk'}$.
The adjoint operator is defined by, {\it e.g.}, (see \cite{wilson} for further discussion)
\beq
(D^\dagger\om)_i=\sum_{j\in<ij>}\om_{ij}~~,~
(D^\dagger\om)_{ijk}=\om_{ij}+\om_{jk}+\om_{ki}~,
\eeq
where the sum is over all the edges $e_{ij}$ connected to a vertex $v_i$ with
a positive sign for edges leaving $v_i$ and negative for edges coming into $v_i$, 
or over all the edges bounding the triangle $f_{ijk}$ with a positive sign if
their orientation is compatible with the orientation of the triangle; $D$ and
$D^\dagger$ are shown graphically in Figure 1.
\begin{figure*}
\centerline{\mbox{\includegraphics[width=4.00in]{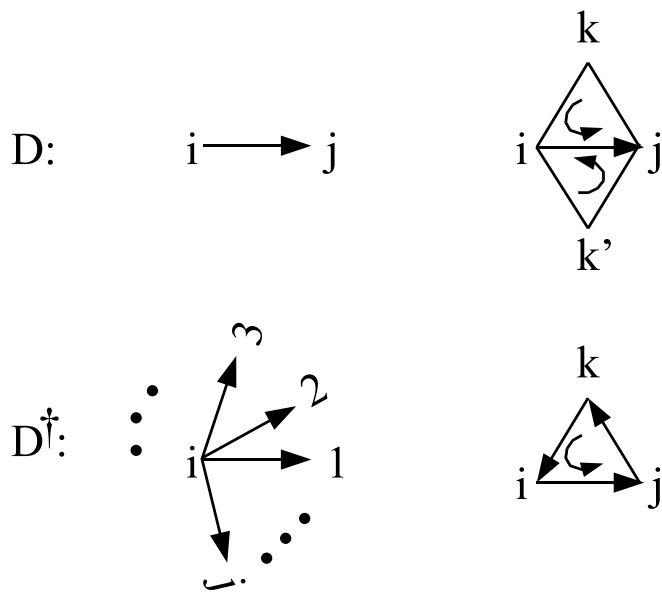}}}
\caption{$D$ and $D^\dagger$}
\end{figure*}
We may now compute the index of this discrete operator $D$; since it maps
$E_0\oplus E_2\to E_1$, it is an $E\times (V+F)$ dimensional matrix,
and the index is simply 
\beq
V+F-E=\chi=2(1-g)~.
\eeq
This is clearly topological, as it does not depend on the values of the entries of
$D$, only on its dimension.

Not only does the Euler character $\chi$ make sense on a 
triangulation of a surface, but its density $\sqrt{g}R$,
where $R(g(x))$ is the scalar curvature\footnote{We use the convention that 
the scalar curvature is minus twice the Gaussian curvature, 
and hence is negative on the sphere \cite{poly}.}
of the two dimensional metric $g_{mn}$, has a
sensible analog as well: since
$\chi= \frac1{2\pi}\sum_{i\in\{V\}}\epsilon_i$ where
$i$ runs over all vertices and $\epsilon_i$ is the defect at the $i$'th vertex, 
one can identify the defect $\epsilon_i=-\frac12\sqrt{g} R_i$ with the 
curvature at the vertex \cite{regge}.
Thus one has the correspondence:
\beq\label{corr}
-\frac1{4\pi}\int d^2x~ \sqrt{g}R=\chi~~\leftrightarrow~~
\frac1{2\pi}\sum_{i\in\{V\}}\epsilon_i=V-E+F~.
\eeq

Anomalies are generally regarded as arising from the infinite numbers of degrees of
freedom in continuous systems. For example, the action $S$ of a scalar field $\Phi$
coupled to a gravitational background on a two dimensional surface $\Sigma$, 
\beq\label{scal}
S = \frac12\int_\Sigma d^2x\sqrt{\det(g_{pq})} g^{mn}\partial_m\Phi\partial_n\Phi~,
\eeq
has a classical symmetry under rescalings of the metric on $\Sigma$:  
$g_{mn}\rightarrow \lambda(x) g_{mn}$. This implies
\beq
g_{mn}\frac{\partial S}{\partial g_{mn}} = 0~.
\eeq
Upon quantization, this symmetry is anomalous; that is, if one defines the quantum effective
action $\Gamma $ as
\beq
e^{-\Gamma[g]}=\int [d\Phi] ~e^{-S[\Phi,g]}~,
\eeq
then one finds \cite{duff,poly}
\beq\label{anom}
g_{mn}\frac{\partial \Gamma}{\partial g_{mn}} = -\frac1{24\pi}\sqrt{\det(g_{mn})}R(g(x))~,
\eeq

If one integrates this over the surface, one finds
\beq
\int d^2x~ g_{mn}\,\frac{\partial \Gamma}{\partial g_{mn}}= -\frac1{24\pi}\int d^2x~ \sqrt{\det(g_{mn})}R(g(x))=\frac16 \chi~.
\eeq

In \cite{kr}, we extended this correspondence 
to the anomaly (\ref{anom}): we found an analog of the effective
action $\Gamma$ on a triangulation. That is, we found a function
$\Gamma(l_{ij})$ of the edge lengths $l_{ij}$ such that
\beq\label{trianom}
\sum_{j\in<ij>} l_{ij} \frac{\partial\Gamma}{\partial l_{ij}} =\frac1{12\pi}\epsilon_i
\eeq
for all vertices $i$ (the sum is over all edges with one end at $i$).

We now present our construction; the remainder of the manuscript
is taken verbatim from \cite{kr}. Our main result is
\beq\label{res}
\Gamma=\frac1{12\pi}\left[\sum_{\angle_{ijk}}\int_{\frac\pi2}^{\alpha_{ijk}}
(y-\frac\pi3)\cot(y)dy +\sum_{<ij>}2k_{ij}\pi\, \ln\!\Big(\frac{l_{ij}}{l_0}\Big)\right],
\eeq
where the first sum is over all internal angles $\alpha_{ijk}$, second sum is over
all edges $<\!ij\!>$ with lengths $l_{ij}$ (the explicit factor of two arises 
because every edge is shared by two triangles), $l_0$ is a scale that we 
set to equal to one from now on, and the $k_{ij}$ are constants associated to
the edges that satisfy
\beq\label{keqs}
\sum_{j\in<ij>}k_{ij}=1-\frac{n_i}6~,\qquad n_i=\sum_{j\in<ij>}1
\eeq
at every vertex $i$; here $n_i$ is the number of neighbors of the $i$'th vertex. Note that the
conditions (\ref{keqs}) do not in general determine the constants $k_{ij}$ uniquely;
one could add a subsidiary condition, {\it e.g.}, that $\sum k_{ij}^2$ is minimized, to 
remove this ambiguity. Note also that the total Euler character, which comes from 
a {\em uniform} scaling of all lengths and thus does not change the angles $\alpha_{ijk}$,
comes entirely from the last term, {\it i.e.}, from ${\partial\Gamma}/{\partial l_0}$.

The strategy that we use to find this solution is as follows: we first consider the simplest case,
a triangulation of the sphere with three vertices, three edges, and two faces,
and prove the integrability conditions needed for $\Gamma$ to exist are satisfied.
We then find $\Gamma$ for this case and show that it immediately generalizes to
all triangulations with a certain homogeneity property, and finally generalize
$\Gamma$ to an arbitrary triangulation.

It would be interesting to complete the correspondence, and 
find a way to compute the result (\ref{res})
as the anomalous effective action corresponding to a discrete analog of, {\it e.g.}, 
the scalar action (\ref{scal}); a promising approach might be the 
work of S.~Wilson on triangulated manifolds \cite{wilson}.

\section{Integrability}
We begin with a triangulation of the sphere with three 
vertices, three edges, and two (identical) faces (a triangular ``pillow''); 
we label the edges by their lengths $a,b,c$
and the opposite internal angles of the triangles by  
$\alpha$, $\beta$ and $\gamma$, respectively. We also abbreviate 
\beq
a\frac{\partial\Gamma}{\partial a}\equiv D_a(\Gamma)~,~etc.,
\eeq
and, for simplicity, drop an overall factor of $1/(12\pi)$ in $\Gamma$. The defect at the
vertex $\alpha$ in this case is just $2(\pi-\alpha)$, {\it etc.}
Using this notation, the equations that we want $\Gamma$ to satisfy in the triangle are:
\bea
D_a(\Gamma)+D_b(\Gamma) &=& 2(\pi - \gamma)~,\cr
D_a(\Gamma)+D_c(\Gamma) &=& 2(\pi - \beta)~,\cr
D_b(\Gamma)+D_c(\Gamma) &=& 2(\pi - \alpha)~.
\eea
If we add the first two equations, and subtract the third, 
we get $ D_a(\Gamma) = (\pi - \gamma - \beta + \alpha)$. 
Since $\gamma + \beta + \alpha$ is $\pi$, we get 
\beq 
D_a(\Gamma) = 2\alpha~,~~D_b(\Gamma) = 2\beta~,~~D_c(\Gamma) = 2\gamma~.
\eeq
The function $\Gamma$ can exist only
if 
\beq\label{integ}
D_b(D_a(\Gamma)) = D_a(D_b(\Gamma))~.
\eeq
Using 
\beq\label{cos}
\alpha=\arccos\left(\frac{-a^2+b^2+c^2}{2bc}\right)~,
\qquad\beta=\arccos\left(\frac{a^2-b^2+c^2}{2ac}\right)~,
\eeq
it is easy to see that (\ref{integ}) is satisfied. Thus the integrability conditions 
are satisfied for the triangle.

The tetrahedron and octahedron give results similar to those of the triangle. However, for
a general triangulation, it is not easy to decouple the integrability conditions
and reduce them to equations that may be checked straightforwardly. Instead, we construct
$\Gamma$ explicitly.

\section{The effective action}
Because the triangle is the simplest case, and can be related to any other system, 
we have examined it in detail. As noted above, in this case the defect at a vertex 
is directly related to the internal angle at that vertex; this suggests that in the 
general case, where the defect is related to the sum of the internal angles at 
a vertex, $\Gamma$ should be just the sum of the $\Gamma$ for each 
triangle. This is almost correct.

The basic strategy for the triangle was to rewrite the differential equations 
for $\Gamma$ in terms of new variables: the angles $\alpha,\beta$, 
and the edge length $c$ between them. One can integrate
some of the equations and finally arrive at $\Gamma$ on a single triangle $\Delta$:
\beq
\Gamma_\Delta = \sum_i\left[ (\alpha_i -\frac\pi3)
\ln(\sin(\alpha_i))-\int_{\frac{\pi}{2}}^{\alpha_i} \ln(\sin(y))dy+k_i\pi \ln(a_i)\right]~,
\eeq
where $\{a_1,a_2,a_3;\alpha_1,\alpha_2,\alpha_3\}=\{a,b,c;\alpha,\beta,\gamma\}$, 
and $k_i$ are constants associated to each edge.
This can be simplified by integration by parts:
\beq
\Gamma_\Delta = \sum_i\left[\int_{\frac\pi2}^{\alpha_i} (y-\frac\pi3)\cot(y)dy
+k_i\pi \ln(a_i)\right]~.
\eeq
Note that all terms are expressed in terms of the internal angles $\alpha_i$ 
except for the last term, which explicitly involves $a_i$. To prove that 
this is correct (and to determine $k_i$), we differentiate $\Gamma$:
\bea
a_i\frac{\partial\Gamma}{\partial a_i}\equiv D_{a_i}\Gamma &=& 
\sum_j\left[(\alpha_j-\frac\pi3)\cot(\alpha_j)D_{a_i}\alpha_j\right]+k_i\pi\cr
& = &\sum_j\left[-(\alpha_j-\frac\pi3)\frac{\cos(\alpha_j)D_{a_i} \cos(\alpha_j)}{1-\cos^2(\alpha_j)}\right]+k_i\pi~.
\eea
Then the contribution of one triangle to the defect at 
vertex $1$ is given by 
\bea
(D_b+D_c)\Gamma&=& -(\alpha-\frac\pi3)\frac{\cos(\alpha)(D_b+
D_c)\cos(\alpha)}{1-\cos^2(\alpha)} -(\beta-\frac\pi3)\frac{\cos(\beta)(D_b+
D_c)\cos(\beta)}{1-\cos^2(\beta)}\nonumber\\  \nonumber\\
&& -(\gamma-\frac\pi3)\frac{\cos(\gamma)(D_b+
D_c)cos(\gamma)}{1-\cos^2(\gamma)}+(k_b+k_c)\pi\nonumber
\eea
which can be explicitly evaluated using $\gamma=\pi-\alpha-\beta$ and (\ref{cos}), and gives:
\beq
(D_b+D_c)\Gamma= \frac\pi3 - \alpha+k_b\pi+k_c\pi~.
\eeq
This calculation works just as well for any triangle in a 
general triangulation to give $\Gamma$ on any surface. 

The constants $k_i$ are assigned to every edge $i$. Clearly, in the case of the 
triangular pillow, there is a trivial solution to $k_i=\frac13$ for all $i$ (recall 
that $\pi-\alpha$ is half the defect at the vertex, but there are two triangles 
meeting at each vertex to sum over in this case). More generally, for any locally
homogeneous triangulation in which all vertices have $n$ nearest neighbors, 
we can choose 
\beq
k_i=\frac1n-\frac16~.
\eeq
However, for a general 
triangulation, the edge may connect vertices with different numbers of 
edges connecting to them. In this case, it is not necessarily trivial to find 
the appropriate values for all the $k_i$.

\section{A Problem in Graph Theory}
On a general triangulated surface, the we have the condition
\beq\label{keq}
\sum_{j\in<ij>}k_{ij}=1-\frac{n_i}6
\eeq
at every vertex, where $n_i$ is the number of neighbors of the $i$'th vertex. 
This means we want to find labels for the edges of a graph such that the sum at
each vertex is the same.
The equations (\ref{keq}) are a system $V$ linear equations on the $E$ variables $k_{ij}$, 
where $V$ is the total number of vertices and $E$ is the total number of edges; 
note that $E\ge V$ for all triangulations. We can rewrite this in terms of the
$V\times E$ dimensional matrix that describes the connections between 
vertices. This matrix has exactly two ones,
which correspond to two vertices, in each column, which corresponds 
to the edge connecting the vertices.

We are happy to thank L.~Motl \cite{motl} for the following proof that
a solution to these equations always exists. The system of equations would not have a
solution only if there were a linear combination of the rows that vanishes, that is, if there
existed a vector that is perpendicular to all the columns. Because every column
contains exactly two ones, it suffices to consider a submatrix that
defines a triangle:
\beq
\left(
\begin{array}{ccc}
1 & 1 & 0\\
1 & 0 & 1\\
0 & 1 & 1\\
\end{array}
\right)~.
\eeq
Since this matrix is nondegenerate,
no nontrivial vector is annihilated by it. 
Since every edge sits on a triangle, no such vector can exist
for the whole triangulation. Therefore, 
there must always be at least one way to label the edges.

\bigskip\bigskip
\noindent{\bf\Large Acknowledgement}:
\bigskip

\noindent We are grateful to the 2005 Simons Workshop
for providing a stimulating atmosphere. MR is grateful to the organizers
of the 26th Winter School GEOMETRY AND PHYSICS at Srni
for the opportunity to present this work.
We are happy to thank Lubo\v s Motl for providing the proof in section 4 
as well as many helpful comments on the manuscript, 
and Ulf Lindstr\"om for his comments.
The work of MR was supported in part by NSF grant no.~PHY-0354776, 
by the University of Amsterdam, and by Stichting FOM.

\end{document}